\documentclass[aps,prl,twocolumn,groupedaddress,showpacs,showkeys]{revtex4-1}

\usepackage{amsmath}
\usepackage{latexsym}
\usepackage{epsfig}
\usepackage{epstopdf}
\usepackage{makeidx}
\usepackage{amsfonts}
\usepackage{mathtools}
\usepackage{color}
\usepackage{float}
\usepackage{appendix}

\begin{document}

\title{ Anomalous Random Neural Networks: a Special Renewal Process}

\author{H. Zhang}
\email{zhanghong13@cdut.cn}
\affiliation{College of Mathematics and Physics, Chengdu University of
Technology, Cheng'du, Si'chuan 610059, China}

\author{G. H. Li}
\email{liguohua13@cdut.cn}
\affiliation{College of Mathematics and Physics, Chengdu University of
Technology, Cheng'du, Si'chuan 610059, China}

\date{\today}

\begin{abstract}
In this paper we propose an open anomalous semi-Markovian random neural networks model with negative and positive signals with arbitrary random waiting times.
We investigate
the signal flow process in the anomalous random neural networks based on renewal process, and obtain the corresponding master equation for time evolution of
the probability of the potential of the neurons. As examples, we discuss the special cases of exponential waiting times and power law ones, and find the fractional memory effect of the probability of the system state on its history evolution. Besides, the closed random neural networks model is introduced  and the corresponding rate equation is given.  

\end{abstract}



\maketitle

\section{Introduction}

Neural network (NN) has been shown to be highly useful in the development of artificial intelligence. \cite{RKHB2012} Random Neural Network (RNN), which has existed since 1989 [2-5],  is a stochastic integer-state integrate
and ﬁre system \cite{YG2016} and  is one of the most important NNs. The RNN has numerous applications in  
deep reinforcement learning \cite{S2022}, generative adversarial networks \cite{S2023}, nonnegative autoencoder \cite{YG2016}, and other fields [9-17].

In RNN, it is simply assumed that the model is with exponential 
signal emission intervals, Poisson external signal arrivals, and Markovian
signal movements between neurons. \cite{G1989} But the problem is more
complicated by the nonlinearity of biological and artiﬁcial
neurons and by the heterogeneity of the system.\cite{CAL2023} In fact, the waiting time for the signal emission and arrivals, et. are generally non-exponential, but obeys heavy-tail distribution. \cite{HTF2013}
 To our knowledge, an analytical framework
 capturing the effect of disorder on signal movement dynamics is still lacking. 
 
 Herein we shall develop an open anomalous semi-Markovian random neural networks model with negative and positive signals with arbitrary random waiting times. 
We investigate
the signal flow process in the anomalous random neural networks based on renewal process, and obtain the corresponding master equation for time evolution of
the probability of the potential of the neurons. The technology challenge to obtain the master equation is to consider the process of signal arriving and leaving, and diffusing as chemical reactions, and to use the renewal method \cite{UIGK2021} in chemical continuous time random walks in Refs.\cite{AD2017}.
As examples, we also discuss the special cases of exponential waiting times and power-law ones, and find that there is a fractional memory effect of the probability of the system state on its history evolution. Finally, the closed random neural networks are studied as an special case and the corresponding rate equation is given.

\section{Classical random neural networks with exponential distributed waiting times }

\subsection{Model}
We first recall the classical random neural networks.
In Ref.\cite{G1989,G1993}, Gelenbe introduced an open random neural networks of $n$ neurons in which "positive" and 
"negative" signals circulate. 
External arrivals of signals to the network 
can either be positive, arriving at the $i-$th neuron according to a Poisson 
process of rate $\Lambda (i)$, or negative according to a Poisson process of rate 
$\lambda (i)$. Positive and negative signals have opposite roles. A negative signal 
reduces by 1 the potential of the neuron to which it arrives (i.e., it "cancels" 
an existing signal) or has no effect if the potential is zero. A positive 
signal adds 1 to the neuron potential. Negative potentials are not allowed 
at neurons. If the potential at a neuron is positive, it may "fire," sending 
signals out toward other neurons or to the outside of the network. As 
signals are sent, they deplete the neuron's potential by the same number. 
The times between successive signal emissions when neuron $i$ fires are 
exponentially distributed random variables of average value $1/ r(i)$; hence 
$r(i)$ is the rate at which neuron $i$ fires. 
A signal leaving neuron i when it "fires" heads for the other neuron $j$ with 
probability $p^{+}_{ij}$ as a positive signal, or as a negative signal with probability
$P^{-}_{ij}$, or departs from the network with probability $d(i)$. Let 
$ P_{ij}= p^{+}_{ij}+ P^{-}_{ij}$; it is the transition probability of a Markov chain 
representing the movement of signals between neurons. It is not allowed that
the signals leaving a neuron to return directly back to the same 
neuron: $P_{ii}=0$ for all $i$. Thus, one has $\sum_{j\neq i}p_{ij}+d_i=1$
Positive signals represent excitation and negative signals represent inhibition.

\subsection{Signal flow process}
The main properties of the model are presented in the following descriptions for signal flow process. \cite{G1989, G1993}

Let $k(t)$ be the vector of neuron potentials at time $t$, and $k =(k_1,...,k_n)$
be a particular value of the vector;  and the component $k_i$ is the potential of the neuron $i$ at time $t$. let $P(k,t)$ denote the probability
distribution $Prob[k(t)=k]$. 
Let $P(k)$ denote the stationary probability distribution
$P(k)=\lim_{t\to \infty}P(k,t)$ if it exists.

Let $q_i$ denote the quantity 
$q_i=\frac{\lambda^{+}_{i}}{r_i+\lambda^{-}_{i}}$,
where the $\lambda^{+}_{i}, \lambda^{-}_{i}$ satisfy the system of nonlinear simultaneous equations:
\begin{equation}
  \lambda^{+}_{i}=\sum_{j\neq i}  q_j r_j p^{+}_{ij}+\Lambda_i,
  \label{lambda1}
\end{equation}
\begin{equation}
  \lambda^{-}_{i}=\sum_{j\neq i}  q_j r_j p^{-}_{ij}+\Lambda_i.
  \label{lambda2}
\end{equation}
If a unique non-negative solution $ \lambda^{+}_{i},\lambda^{-}_{i}$ exists to equations $\ref{lambda1}$,  $\ref{lambda2}$ such that $q_i<1$, then
\begin{equation}
    P(k)=(1-q_i)q^{k_i}_j.
\end{equation}
Since $\{k(t): t \geq 0\}$ is a continuous time Markov chain it satisfies the usual Chapman-Kolmogorov equations, thus in steady-state it can be seen that $p(k)$ must satisfy the following global balance equations:
\begin{eqnarray}
&&\sum_{i=1}^{m}P(k)[\Lambda_i+(\lambda_i+r_i)1[k_i>0]]\nonumber\\
&&=\sum_{i=1}^{m}\sum_{j\neq i}^{m}
[P(k-s_{\Lambda_i})\Lambda_i 1[k_i>0]\nonumber\\
&&+P(k-s_{\lambda_i})\lambda_i\nonumber\\
&&+P(k-s_{r_i}^{+})r_i P^{+}_{ij}1[k_j>0]\nonumber\\
&&+P(k-s_{r_i}^{-})r_i P^{-}_{ij}\nonumber\\
&&+P(k-s_{r_i b}^{-})r_i P^{-}_{ij}1[k_j=0]\nonumber\\
&&+P(k-s_{r_i b}^{-})r_i d_i].
\end{eqnarray}
where the  vectors for the potentials of changing are
$s_{\Lambda i}=\binom{0,...,1,...,0}{  i}$ for the positive signal arriving,
$s_{\lambda i}=\binom{0,...,-1,...,0}{  i}$ for the negative signal arriving,
$s_{r_i }^{+}=\binom{0,...,-1,...,1,...,0}{  i,...,j}$ for the positive signal transition from $i$ to $j$,
$s_{r_i }^{-}=\binom{0,...,-1,...,-1,...,0}{  i,...,j}$ for the negative signal transition from $i$ to $j$,
and $s_{r_i b}^{-}=\binom{0,...,-1,...,0}{  i},$ for the signal departing from the network.  And $I[X]$ is the usual characteristic function which takes the  value 1 if $X$ is true and $0$ otherwise.

\section{Anomalous random neural networks with arbitrarily distributed waiting times}

\subsection{Model}

We now develop an anomalous random neural networks model
 where each neuron in the system can receive negative and positive with respective random waiting time from outside, and the neuron with positive potential has to wait some time to fire and then transfer the negative or positive signal to the other neuron or the outside of the system.

The difference of this model with classical one is that the arriving process is not according to a Poisson 
process. We assume that  
the waiting time $t_1$ between positive signal arriving neuron $i$ from outside, the waiting time $t_2$ between negative signal arriving neuron $i$ from outside, and the waiting time $t_3$  between the neuron $i$ firing are
independent random variables, and the probability density function (PDFs) of the waiting times  are  
$\psi_{\Lambda i}(t_1)$, $\psi_{\lambda i}(t_2)$ and $\tilde{\psi}_{r_i}(t_3)$, respectively, which are arbitrarily distributed and not needed to be exponential.

 One can see that the waiting time PDF for the occurrence of the neuron $i$ successful firing is in fact $\psi_{ri}(t_3)=\tilde{\psi}_{ri}(t_3)1[k_i>0]$ because of the positive condition. Hereafter, we shall use $\psi_{r_i}(t_3)$ denote the waiting time PDF for the neuron $i$ firing for simplicity.

\subsection{Signal flow process with arbitrarily distributed firing and arriving waiting times}

Next we will consider the signal flow process in the anomalous random neural networks.

Let $\tau=\min\{\tau_{\Lambda 1},\tau_{\lambda 1},\tau_{r1},...,\tau_{\Lambda m},\tau_{\lambda m},\tau_{rm}\}$ be the minimum waiting time
of external arrivals of signals and firing for all neurons. Let $\phi_{\Lambda i}(\tau)$, $\phi_{\lambda i}(\tau)$ and $\phi_{ri}(\tau)$ denote the pdfs of the minimum waiting time $\tau$ being the arriving of positive signal, negative signal and the occurring of firing of neuron $i$  respectively. Here the minimum waiting times  are assumed to be independent on the potential $k$.
Then we have
\begin{eqnarray}
\phi_{\Lambda i}(t)&=&\psi_{\Lambda i}(t)\Psi_{\lambda i}(t)\Psi_{ri}(t)\prod_{l\neq i}^{m}\Psi_{\Lambda j}(t)\nonumber\\
&&\cdot\Psi_{\lambda j}(t)\Psi_{rj}(t)
\end{eqnarray}
\begin{eqnarray}
\phi_{\lambda i}(t)&=&\psi_{\lambda i}(t)\Psi_{\Lambda i}(t)\Psi_{ri}(t)\prod_{l\neq i}^{m}\Psi_{\Lambda j}(t)\nonumber\\
&&\cdot\Psi_{\lambda j}(t)\Psi_{rj}(t)
\end{eqnarray}
\begin{eqnarray}
\phi_{r_i}(t)&=&\tilde{\psi}_{r_i}(t)1[k_i>0]\Psi_{\Lambda i}(t)\Psi_{\lambda i}(t)\nonumber\\
&&\cdot\prod_{l\neq i}^{m}\Psi_{\Lambda j}(t)\Psi_{\lambda j}(t)\Psi_{r_j}(t)
\label{rwaiting}
\end{eqnarray}
where $\Psi_{\Lambda i}(t)=1-\int_{0}^{t}\psi_{\Lambda i}(t')dt'$,
$\Psi_{\lambda i}(t)=1-\int_{0}^{t}\psi_{\lambda i}(t')dt'$,
and
$\Psi_{r_i}(t)=1-\int_{0}^{t}\tilde{\psi}_{r_i}(t')1[k_i>0]dt'$.
Let $\Phi(t)$ be the probability without any external arrivals of signals or firing of neurons in the time interval $[0,t]$, that is,
\begin{eqnarray}
\Phi(t)=\prod_{i}^{m}\Psi_{\Lambda i}(t)\Psi_{\lambda i}(t)\Psi_{r_i}(t).
\label{survival time}
\end{eqnarray}
One can find that
\begin{eqnarray}
-\frac{\partial\Phi(t)}{\partial t}&=&\sum_{i=1}^{m}[\phi_{\Lambda i}(t)\nonumber\\
&&+\phi_{\lambda i}(t)+\phi_{r_i}(t)],
\end{eqnarray}
which means
\begin{eqnarray}
\Phi(t)&=&\int_{t}^{\infty}\sum_{i=1}^{m}[\phi_{\Lambda i}(t')\nonumber\\
&&+\phi_{\lambda i}(t')+\phi_{r_i}(t')]dt',
\end{eqnarray}
and then
\begin{eqnarray}
\Phi(u)&=&\frac{1}{u}\{1-\sum_{i=1}^{m}[\phi_{\Lambda i}(u)\nonumber\\
&&+\phi_{\lambda i}(u)+\phi_{r_i}(u))]\},
\end{eqnarray}
where $\Phi(u)$, $\phi_{\Lambda i}(u)$, $\phi_{\lambda i}(u)$ and $\phi_{r_i}(u)$ are the Laplace
$t\rightarrow u$ transforming of $\Phi(t)$, $\phi_{\Lambda i}(t)$, $\phi_{\lambda i}(t)$ and $\phi_{r_i}(t)$ respectively.

Let us consider the recursion relations
\begin{eqnarray}
k_{n+1}=k_n+s_n,
\end{eqnarray}
\begin{eqnarray}
T_{n+1}=T_n+\tau_n,
\end{eqnarray}
for the potential vector $K_n$ and time $T_k$ describe an stochastic process  after $n$ steps \cite{AD2017}.
One can see that the probability $P(k,t)$ for the potential $k$ at time $t$ for the neurons can be
described by the probability arriving at $k$ after $n$ steps without any steps from then on, that is,
\begin{eqnarray}
P(k,t)=\int_{0}^{t}dt'\sum_{n=0}^{\infty}R_n(k,t')\Phi(t-t').
\end{eqnarray}
Here,
\begin{eqnarray}
R_{n+1}(k,t)&=&\int_{0}^{t}dt'\sum_{i=1}^{m}\sum_{j\neq i}^{m}[R_n(k-s_{\Lambda i},t')\nonumber\\
&&\cdot\phi_{\Lambda i}(t-t')1[k_i>0]\nonumber\\
&&+R_n(k-s_{\lambda i},t')\phi_{\lambda i}(t-t')\nonumber\\
&&+R_n(k,t')\phi_{\lambda i}(t-t')1[k_i=0]\nonumber\\
&&+R_n(k-s_{r_i}^{+},t')\phi_{r_i}(t-t')\nonumber\\
&&\cdot P^{+}_{ij}1[k_j>0]\nonumber\\
&&+R_n(k-s_{r_i}^{-},t')\phi_{r_i}(t-t')P^{-}_{ij}\nonumber\\
&&+R_n(k-s_{r_i b}^{-},t')\phi_{r_i}(t-t')\nonumber\\
&&\cdot P^{-}_{ij}1[k_j=0]\nonumber\\
&&+R_n(k-s_{r_i b}^{-},t')\phi_{r_i}(t-t')d_i,
\end{eqnarray}

Besides, $R_{n}(k,t)$ is probability
distribution of just arriving at $k$ at time $t$ after
$n$ steps, and for $R_0(k,t)$ one has
\begin{eqnarray}
R_{0}(k,t)=P(k,0)\delta(t).
\end{eqnarray}

Let $R(k,t)=\sum_{n=0}^{\infty}R_n(k,t)$. If $R(k,u), R_n(k,u)$ are respectively the Laplace transform of $R(k,t), R_n(k,t)$, then we have
\begin{eqnarray}
R(k,u)&=&\sum_{n=0}^{\infty}R_n(k,u)\nonumber\\
&&\sum_{i=1}^{m}\sum_{j\neq i}^{m}[R(k-s_{\Lambda i},u)\nonumber\\
&&\phi_{\Lambda i}(u)1[k_i>0]\nonumber\\
&&
+R(k-s_{\lambda i},u)\phi_{\lambda i}(u)\nonumber\\
&&+R(k,u)\phi_{\lambda i}(u)1[k_i=0]\nonumber\\
&&+R(k-s_{r_i}^{+},u)\phi_{ri}(u)\nonumber\\
&&\cdot P^{+}_{ij}1[k_j>0]\nonumber\\
&&+R(k-s_{r_i}^{-},u)\phi_{ri}(u) P^{-}_{ij}\nonumber\\
&&+R(k-s_{r_i b}^{-},u)\phi_{ri}(u)\nonumber\\
&&\cdot P^{-}_{ij}1[k_j=0]\nonumber\\
&&+R(k-s_{r_i b}^{-},u)\phi_{r_i}(u)d_i\nonumber\\
&&+P(k,0).
\end{eqnarray}
Obviously,
\begin{eqnarray}
P(k,u)&=&\sum_{n=0}^{\infty}R_n(k,u)\Phi(u)=R(k,u)\Phi(u),
\end{eqnarray}
and so
\begin{eqnarray}
R(k,u)=P(k,u)\frac{1}{\Phi(u)}.
\end{eqnarray}
Besides, from (18) we find
\begin{eqnarray}
P(k,u)&=&R(k,u)\frac{1}{u}\{1-\sum_{i=1}^{m}[\phi_{\Lambda i}(u)\nonumber\\
&&+\phi_{\lambda i}(u)+\phi_{r i}(u))]\},
\end{eqnarray}
Combining Eqs. (17), (19) and (20), we find the master equation for the signal flow process in anomalous neural network in Laplace space:
\begin{eqnarray}
&&uP(k,u)-P(k,0)=[R(k,u)-P(k,0)]-\sum_{i=1}^{m}R(k,u)\nonumber\\
&&\cdot[\phi_{\Lambda i}(u)+\phi_{\Lambda i}(u)+\phi_{r i}(u)]\nonumber\\
&&=\sum_{i=1}^{m}\sum_{j\neq i}^{m}
[P(k-s_{\Lambda i},u)\frac{\phi_{\Lambda i}(u)}{\Phi(u)}1[k_i>0]\nonumber\\
&&
+P(k-s_{\lambda i},u)\frac{\phi_{\lambda i}(u)}{\Phi(u)}\nonumber\\
&&+P(k,u)\frac{\phi_{\lambda i}(u)}{\Phi(u)}1[k_i=0]\nonumber\\
&&+P(k-s_{r_i}^{+},u)\frac{\phi_{ri}(u)}{\Phi(u)}P^{+}_{ij}1[k_j>0]\nonumber\\
&&+P(k-s_{r_i}^{-},u)\frac{\phi_{ri}(u)}{\Phi(u,k)}P^{-}_{ij}\nonumber\\
&&+P(k-s_{r_i b}^{-},u)\frac{\phi_{ri}(u)}{\Phi(u)}P^{-}_{ij}1[k_j=0]\nonumber\\
&&+P(k-s_{r_i b}^{-},u)\frac{\phi_{ri}(u)}{\Phi(u,k)}d_i]\nonumber\\
&&-\sum_{i=1}^{m}P(k,u)\frac{1}{\Phi(k,u)}[\phi_{\Lambda i}(u)\nonumber\\
&&+\phi_{\lambda i}(u)+\phi_{ri}(u)].
\end{eqnarray}
Inverting Eq.(21) to time space yields the master equation in time space
\begin{eqnarray}
&&\frac{\partial P(k,t)}{\partial t}=\sum_{i=1}^{m}\sum_{j\neq i}^{m}
\int_{0}^{t}[P(k-s_{\Lambda i},t')\nonumber\\
&&\Theta_{\Lambda i}(t-t')1[k_i>0]\nonumber\\
&&
+P(k-s_{\lambda i},t')\Theta_{\lambda i}(t-t')\nonumber\\
&&
+P(k-s_{r_i}^{+},t')\Theta_{ri}(t-t')P^{+}_{ij}1[k_j>0]\nonumber\\
&&+P(k-s_{r_i}^{-},t')\Theta_{ri}(t-t') P^{-}_{ij}\nonumber\\
&&+P(k-s_{r_i b}^{-},t')\Theta_{ri}(t-t') P^{-}_{ij}1[k_j=0]\nonumber\\
&&+P(k-s_{r_i b}^{-},t')\Theta_{ri}(t-t')d_i]dt'\nonumber\\
&&-\sum_{i=1}^{m}\int_{0}^{t}P(k,t')[\Theta_{\Lambda i}(t-t')\nonumber\\
&&+\Theta_{\lambda i}(t-t')1[k_i>0]+\Theta_{r i}(t-t')]dt',
\end{eqnarray}
where $\Theta_{\Lambda i}(t),\Theta_{\lambda i}(t)$ and $\Theta_{r i}(t)$ are respectively the inverse Laplace transforms $u\rightarrow t$ of
$$\Theta_{\Lambda i}(u)=\frac{\phi_{\Lambda i}(u)}{\Phi(u)},$$
$$\Theta_{\lambda i}(u)=\frac{\phi_{\lambda i}(u)}{\Phi(u)},$$
and
$$\Theta_{r_i}(u)=\frac{\phi_{ri}(u)}{\Phi(u)}.$$
In the steady state we get
\begin{eqnarray}
&&\sum_{i=1}^{m}\sum_{j\neq i}^{m}
\int_{0}^{t}[P(k-s_{\Lambda i},t')\nonumber\\
&&\Theta_{\Lambda i}(t-t')1[k_i>0]\nonumber\\
&&
+P(k-s_{\lambda i},t')\Theta_{\lambda i}(t-t')\nonumber\\
&&+P(k-s_{r_i}^{+},t')\Theta_{ri}(t-t')P^{+}_{ij}1[k_j>0]\nonumber\\
&&+P(k-s_{r_i}^{-},t')\Theta_{ri}(t-t') P^{-}_{ij}\nonumber\\
&&+P(k-s_{r_i b}^{-},t')\Theta_{ri}(t-t') P^{-}_{ij}1[k_j=0]\nonumber\\
&&+P(k-s_{r_i b}^{-},t')\Theta_{ri}(t-t')d_i]dt'\nonumber\\
&&=\sum_{i=1}^{m}\int_{0}^{t}P(k,t')[\Theta_{\Lambda i}(t-t')\nonumber\\
&&+\Theta_{\lambda i}(t-t')1[k_i>0]+\Theta_{r_i}(t-t')]dt'.
\end{eqnarray}

\section{Examples}
\subsection{Exponential case}

Let us consider the special case for exponential waiting time $\psi_{\Lambda i}(t)=\Lambda_i e^{-\Lambda_i t}$,
$\psi_{\lambda i}(t)=\lambda_i e^{-\lambda_i t}$ and $\psi_{ri}(t)=r_i e^{-r_i t}1[k_i>0]$.
In this case we find
$\Psi_{\Lambda i}(t)=e^{-\Lambda_i t}$,
$\Psi_{\lambda i}(t)= e^{-\lambda_i t}$
and $\Psi_{r_i}(t)=(e^{-r_i t}-1)1[k_i>0]+1$,
and then
$$\Phi(t)=\exp(-\sum_{i=1}^{m}\Lambda_i t)\exp(-\sum_{i=1}^{m}\lambda_i t)\exp(-\sum_{i=1}^{m}r_i1[k_i>0] t),$$
$$\phi_{\Lambda i}(t)=\Lambda_i \Phi(t),$$
$$\phi_{\lambda i}(t)= \lambda_i \Phi(t),$$
$$\phi_{ri}(t)=ri 1[k_i>0]\Phi(t).$$
In the Laplace space we get
$\Theta_{\Lambda i}(u)=\Lambda i,$
$\Theta_{\lambda i}(u)=\lambda i,$
and
$\Theta_{r_i}(u)=r_i 1[k_i>0],$
which yields
$$\Theta_{\Lambda i}(t)=\Lambda i\delta(t),$$
$$\Theta_{\lambda i}(t)=\lambda i\delta(t),$$
and
$$\Theta_{r_i}(t)=r_i 1[k_i>0]\delta(t).$$
We substitute them into the steady equation and recover the global balance equation in Ref.[1].

\subsection{Power-law firing waiting time for single neuron }

Let us consider the other special case where all the assumptions are the same except the firing waiting time of a single neuron $i$. We assume  $\psi_{\Lambda i}(t)=\Lambda_i e^{-\Lambda_i t}$,
$\psi_{\lambda i}(t)=\lambda_i e^{-\lambda_i t}$, $\psi_{ri}(t)=\frac{\tau_0^{\beta}}{\Gamma(1-\beta)}\frac{1}{t^{1+\beta}}1[k_i>0]$ for $0<\beta<1$, and $\psi_{rl}(t)=r_l e^{-r_l t}1[k_l>0]$ for $l\neq i$.
Thus,
 $\Psi_{ri}(t)=\frac{\tau_0^{\beta}}{\Gamma(1-\beta)}\frac{1}{t^{\beta}}1[k_i>0]+1[k_i\leq 0]$,
and then
\begin{eqnarray}
 \Phi(t)&&=\exp(-\sum_{i=1}^{m}\Lambda_i t)\exp(-\sum_{i=1}^{m}\lambda_i t)\nonumber\\
   &&\cdot\exp(-\sum_{j\neq i}^{m}r_j 1[k_j >0] t)\nonumber\\
   &&\cdot\bigg[\frac{\tau_0^{\beta}}{\Gamma(1-\beta)}\frac{1}{t^{\beta}}1[k_i >0]+1[k_i\leq 0]\bigg].    
\end{eqnarray}
We also get 
$$\phi_{\Lambda i}(t)=\Lambda_i \Phi(t),$$
$$\phi_{\lambda i}(t)= \lambda_i \Phi(t),$$
$$\phi_{rl}(t)=r_l 1[k_l>0]\Phi(t),$$
and
\begin{eqnarray}
 \phi_i(t)&&=\frac{\tau_0^{\beta}}{\Gamma(1-\beta)}\frac{1}{t^{1+\beta}}1[k_i>0]\nonumber\\
 &&\cdot\exp(-\sum_{i=1}^{m}\Lambda_i t)\exp(-\sum_{i=1}^{m}\lambda_i t)\nonumber\\
   &&\cdot\exp(-\sum_{l\neq i}^{m}r_l 1[k_l >0] t).      
\end{eqnarray}
In the Laplace space we find
$\Theta_{\Lambda i}(u)=\Lambda i,$
$\Theta_{\lambda i}(u)=\lambda i,$
$\Theta_{r_l}(u)=r_l 1[k_l>0],$ and 
$\Theta_{r_i}(u)=\frac{1}{\tau_0^{\beta}}(u+\alpha)^{1-\beta}1[k_i>0],$
where $\alpha=\sum_{i=1}^{m}\Lambda_i+\sum_{i=1}^{m}\lambda_i+\sum_{l\neq i}^{m}r_l 1[k_l >0]$.
We substitute them into the equation (21) and take the inverse Laplace transform and obtain the fractional evolution equation
\begin{eqnarray}
&&\frac{\partial P(k,t)}{\partial t}=\sum_{i=1}^{m}
\bigg[P(k-s_{\Lambda_i},t)\Lambda_i 1[k_i>0]\nonumber\\
&&+P(k-s_{\lambda_i},t)\lambda_i\bigg]\nonumber\\
&&+\sum_{l\neq i}^{m}\sum_{j\neq l}^{m}\bigg[
P(k-s_{r_l}^{+},t)r_l P^{+}_{lj}1[k_j>0]\nonumber\\
&&+P(k-s_{r_l}^{-},t)r_l P^{-}_{lj}\nonumber\\
&&+P(k-s_{r_l b}^{-},t)r_l P^{-}_{lj}1[k_j=0]\nonumber\\
&&+P(k-s_{r_l b}^{-},t)r_ld_i\bigg]\nonumber\\
&&
+\frac{1}{\tau_0^{\beta}}\bigg\{e^{-\alpha t} D_t^{1-\beta}[e^{\alpha t}P(k-s_{r_i}^{+},t)]P^{+}_{ij}1[k_j>0]\nonumber\\
&&+e^{-\alpha t} D_t^{1-\beta}[e^{\alpha t}P(k-s_{r_i}^{-},t)] P^{-}_{ij}\nonumber\\
&&+e^{-\alpha t} D_t^{1-\beta}[e^{\alpha t}P(k-s_{r_i b}^{-},t)] P^{-}_{ij}1[k_j=0]\nonumber\\
&&+e^{-\alpha t} D_t^{1-\beta}[e^{\alpha t}P(k-s_{r_i b}^{-},t)]d_l\bigg\}\nonumber\\
&&-\sum_{ i=1}^{m}\big[P(k,t)(\Lambda_i+\lambda_i 1[k_i>0])\big]\nonumber\\
&&-\frac{1}{\tau_0^{\beta}}e^{-\alpha t} D_t^{1-\beta}[e^{\alpha t}P(k,t)]1[k_i>0]\nonumber\\
&&-\sum_{ l\neq i}^{m} P(k,t)r_l1[k_l>0].
\end{eqnarray}
Here, $ D_t^{1-\beta}$ is the Riemann-Liouville fractional derivative operator. This means that the time evolution has fractional memory of its history.

\section{Rate equation}
\subsection{Model for closed random neural networks}

In this special case we consider the system where there is no negative and positive  signal coming from outside, nor going away from inside, and the neuron with positive potential waits some time to fire and then transfer only positive signal to the other.  One can see that the sum of potential of the system is invariable, so we call such system closed neural networks system. 

Let $\tau=\min\{\tau_{r 1},...,\tau_{r m}\}$ be the minimum firing waiting time of  all neurons. 
Thus, Eqs.(\ref{rwaiting}) and (\ref{survival time}) becomes
\begin{eqnarray}
\phi_{r_i}(t)&=&\tilde{\psi}_{r_i}(t)1[k_i>0]\prod_{l\neq i}^{m}\Psi_{r_j}(t)
\end{eqnarray}
and 
\begin{eqnarray}
\Phi(t)=\prod_{i}^{m}\Psi_{r_i}(t),
\end{eqnarray}
and then
\begin{eqnarray}
\Phi(u)&=&\frac{1}{u}\{1-\sum_{i=1}^{m}\phi_{r_i}(u))\}.
\end{eqnarray}

Using the approach of renewal before, one can get the master equation for closed neural networks as following
\begin{eqnarray}
&&\frac{\partial P(k,t)}{\partial t}=\sum_{i=1}^{m}\sum_{j\neq i}^{m}
\int_{0}^{t}[P(k-s_{r_i}^{+},t')\Theta_{ri}(t-t')P^{+}_{ij}dt'\nonumber\\
&&-\sum_{i=1}^{m}\int_{0}^{t}P(k,t')\Theta_{r i}(t-t')dt'.
\end{eqnarray}

\subsection{Rate equation}
If there is only one  positive potential at the initial time, we want to know the rate equation for the probability $P_l(t)$ for neuron $i$ having this positive potential at time $t$. Let $P_l(t)=\sum_{k: k_l=1}P(k, t)$, we get
\begin{eqnarray}
&&\frac{\partial \sum_{k: k_l=1} P(k,t)}{\partial t}\nonumber\\
&&=\sum_{i=1}^{m}\sum_{j\neq i}^{m}\int_{0}^{t}\sum_{k: k_l=1}P(k-s_{r_i}^{+},t')\Theta_{ri}(t-t')P^{+}_{ij}dt'\nonumber\\
&&-\sum_{i=1}^{m}\int_{0}^{t} (\sum_{k: k_l=1}P(k,t'))\Theta_{r i}(t-t')dt'.
\label{closedsystem}
\end{eqnarray}
Since  only a neuron has non-zero potential at each time, and then for the $k$ whose $l-$th component is $1$, others are $0$,    $s_{r_i}^{+}$ must be the vector whose $l-$th component is $1$, the other component $j\neq i$ is $-1$, and the left ones are all $0$, or there are two positive potentials in $k-s_{r_i}^{+}$. Therefore,
we have $k-s_{r_i}^{+}=\binom{0,...,1,...,0}{  j}$, so Eq.(\ref{closedsystem}) becomes
\begin{eqnarray}
&&\frac{\partial P_l(t)}{\partial t}=
\sum_{j\neq i}^{m}\int_{0}^{t}P_j(t')\Theta_{rj}(t-t')P^{+}_{jl}dt'\nonumber\\
&&-\int_{0}^{t} P_l(t')\Theta_{r l}(t-t')dt'.
\end{eqnarray}
Here,  $P_j(t)=\sum_{k: k_j=1}P(k, t)$. This is the rate equation for the closed neural networks.

\section{Conclusion}
In conclusion, in order to describe the signal movement in disordered biological and artiﬁcial
neurons networks, we introduce the anomalous RNN, in which the waiting times for firing are arbitrary distributed, based on RNN proposed by Gelenbe where the distribution of waiting times for firing are exponential for homogeneous neurons networks. By using renewal theory we derive the master equation (22) for the time evolution of the potentials of the neurons. As examples, we respectively discuss the cases for exponential and power law waiting times, in the exponential case we recover the balance equation of RNN, in the power law case we find a fractional memory effect of the distribution of the potential. Finally, we investigate the closed RNN, and get the rate equation (32). This study will provide a new insight in the development of RNN, and can be applied in deep learning, Boltzmann machine, and other machine learning algorithms. 

\section{acknowledgments}
 This work was supported by the Natural Science Foundation of Sichuan Province (Grant
No. 2022NSFSC1752).\newline


\begin{thebibliography}{}

\bibitem[1]{RKHB2012}
L. M. Raff, R. Komanduri,
M. Hagan, and S. T.S. Bukkapatnam, {\it Neural Networks in Chemical Reaction Dynamics}, Oxford University Press (2012)
\bibitem[2] {G1989}
E. Gelenbe,  {\it Neural Computation} {\bf  1}, 502 {1989}
\bibitem[3]{G1990}
E. Gelenbe,  {\it Neural Computation} {\bf  2}, 239 {1990}
\bibitem[4] {G1993}
E. Gelenbe,  {\it Elektrik} {\bf  1}, 27 {1993}
\bibitem[5] {G1993-2}
E. Gelenbe,  {\it Neural Computation} {\bf  5}, 154 {1993}
\bibitem [6] {YG2016}
Y. H. Yin, and E. Gelenbe, {\it International Joint Conference on Neural Networks}, 1 (2016)
\bibitem[7]{S2022}
W. Serrano, {\it Engineering Applications of Artificial Intelligence 
 } {\bf 110},  104751 (2022)
\bibitem[8]{S2023}
W. Serrano, {\it Neural Networks } {\bf  165},  420  (2023)
\bibitem[9]{GKP1993}
E. Gelenbe, V. Koubi, and F. Pekergin,  {\it Proceedings of IEEE Systems Man and Cybernetics Conference - SMC} {\bf 2}, 630 (1993)
\bibitem[10]{CG2000}
 C. E. Cramer and E. Gelenbe, {\it IEEE Journal on Selected Areas in Communications} {\bf 18},  150 (2000)
 \bibitem[11]{GH2002}
E. Gelenbe and K. F. Hussain, {\it IEEE Transactions on Neural Networks} {\bf 13}, 1257  (2002)
\bibitem[12]{G2009}
E. Gelenbe, {\it Communications of the ACM} {\bf 52}, 66  (2009)
\bibitem[13]{GW2012}
E. Gelenbe and F.J. Wu, {\it Computers
and Mathematics with Applications}  {\bf 64}, 3869  (2012)
\bibitem[14]{RV2004}
G. Rubino and M. Varela, {\it First International Conference on the Quantitative Evaluation of Systems}, 110 (2004)
\bibitem[15]{MR2002}
S. Mohamed and G. Rubino, {\it IEEE Transactions on Circuits and Systems for Video Technology} {\bf 12} 1071 (2002)
\bibitem[16]{RL2011}
K. Radhakrishnan and H. Larijani, {\it Performance Evaluation} {\bf 68}, 347 (2011)
\bibitem[17]{GL2014}
T. Ghalut and H. Larijani, {\it Communication Systems, Networks and Digital Signal Processing},  9th International Symposium on. IEEE, 519 (2014)
\bibitem[18]{CAL2023}
D. G. Clark, L. F. Abbott, and A. Litwin-Kumar, arXiv:2207.12373v3[q-bio.NC](2023)
\bibitem [19] {CQG2022}
 J.Z. Chen, C.K. Qu, and P.L. Gong {\it Neural Networks } {\bf 149}, 18 (2022)
\bibitem[20]{HTF2013}
N. Hiratani, J. Teramae, and T. Fukai, {\it Frontiers in computational neuroscience} {\bf 6}, 102 (2013)
 \bibitem[21]{UIGK2021}
 S. Unicomb, G. Iñiguez, J. P. Gleeson, and M. Karsai, {\it Nature communications} {\bf 12} 1 (2021) 
 \bibitem[22] {AD2017}
T. Aquino, and M. Dentz,  {\it  Phys. Rev. Lett.} {\bf 119}, 230601, {2017}




\end{thebibliography}
\end{document}